\documentclass[aps,twocolumn,groupedaddress, showpacs,prb]{revtex4}
\usepackage{graphicx}%

\begin{document}
\bibliographystyle{apsrev}


\title{Thermal roots of correlation-based complexity}


\author{P. Fraundorf}
\affiliation{Physics \& Astronomy/Center for Molecular Electronics, U. Missouri-StL (63121), St. Louis, MO, USA}
\email[]{pfraundorf@umsl.edu}


\date{\today}

\begin{abstract}

Bayesian maxent lets one integrate thermal physics and information 
theory points of view in the quantitative study of complex systems.  
Since net surprisal (a free energy analog for measuring 
``departures from expected'') allows one to place second law 
constraints on mutual information (a multi-moment measure 
of correlations), it makes a quantitative case for the role of 
reversible thermalization in the natural history of invention, and 
suggests multiscale strategies to monitor standing crop as well.  
It prompts one to track evolved complexity starting from live
astrophysically-observed processes, rather than only from
evidence of past events.  Various gradients and boundaries 
that play a role in availability flow, ranging from the edge of 
a wave-packet to the boundary between idea-pools, allow one to 
frame wide-ranging correlations (including that between a 
phenomenon and its explanation) as delocalized {\em physical} structures.  

\end{abstract}
\maketitle

\tableofcontents

\section{Introduction}
\label{sec:Intro}

Eric Chaisson in this journal recently discussed 
the evolution of complex systems, and its empirically-observed 
correlation with free-energy density\cite{Chaisson04}.  This paper is partly 
about the need he cites for putting such energy flows into an agreeable
information-theory context.  However ``perched on the dawn'' 
is not the whole story, since humans also face a decline 
in one of their major sources of free-energy\cite{Musser05}.  
Thus as we (for reasons that have nothing to do with fossil 
fuels\cite{Ward03}) move past the prime of earth's
``age of plants and animals'', a quantitative look at what we want 
to protect about evolved complexity is timely as well.

Historically, information theory since the days of Shannon\cite{ShannonWeaver49} 
and Jaynes\cite{Jaynes57a, Jaynes57b} saw entropy and ensembles as tools 
for applying gambling theory (statistical inference) to 
physical systems with large numbers of similar and/or identical constituents. 
This paradigm\cite{Kuhn70} has worked its way into many 
advanced\cite{Grandy87, Plischke89, Garrod95} and senior
undergraduate\cite{Reif65, Katz67, Girifalco73, Kittel80p, Stowe84p, 
Baierlein99, Schroeder00} textbooks on
statistical physics. However, in spite of the growing application
of these tools in biological and computer sciences, a solid 
`interdisciplinary umbrella' for relating energy and information
is still needed\cite{Marijuan96, Chaisson04}.  
Even the simplifications\cite{Castle65} that 
it affords to the introductory physics student (with few 
exceptions \cite{Moore98}) are not yet available 
in texts. 

The objective in this article is to remind readers of the 
physical context for a Bayesian view of correlations in 
complex systems, and to suggest integrated ways to work
toward multiscale quantitation.  The target audience 
is complex systems researchers in varied fields, as well as students 
in the code-based sciences.  Hence we'll start slowly, but 
after the opening section will point to the literature where 
possible for technical specifics.  

\section{Augmenting the sum of parts}
\label{sec:Augmenting}

Let's begin with correlation itself, i.e. with 
quantitative ways of seeing the whole as more than 
the sum of its parts.  Statistics courses often focus 
on 2nd-moment pair, or variance-based, measures 
like correlation-coefficient and covariance.  
The focus here instead is on logarithmic
(e.g. bitwise) measures of correlation, 
like mutual information, which {\em a priori} 
at least operate on all scales.  

In the context of general system theory\cite{Bertalannfy68}, 
begin by defining two subsystems A and B  
e.g. as individual 
particles, as collections of particles, as individual 
states (which may or may not be occupied with 
particles), or as regions or control volumes in and out 
of which energy and mass might flow, etc. 
Mutual information is defined as what you 
learn about A by knowing B, and vice versa. 

In mathematical terms, one can say for 
subsystems $A$ and $B$ that mutual information 
\begin{equation}
M[A|B] \equiv S_A + S_B - S_{AB},
\label{IAB}
\end{equation} 
where $S_A$ and 
$S_B$ are uncertainties associated with each system 
taken alone, while $S_{AB}$ is uncertainty about 
the combination given all available information, 
including that associated with correlations.  It's 
relatively easy \cite{Lloyd89b} to prove that $M[A|B] \geq 0$.  
In a sense, therefore, $M[A|B]$ is a quantitative 
measure of how systems A and B taken together 
may be more than the sum of their parts.

For example, 
imagine that you have two drawers (A and B) 
for your socks.  Suppose you know that each 
drawer contains $N$ socks, and that the 
socks are identical except they are either 
black or white with equal probability.  
The amount of uncertainty (average surprisal) 
about the socks in A and B given this 
information is $2N$ bits, i.e. $k \ln [2]$ 
or one bit for each sock.  However, if you 
also know that socks were put into the drawers by 
breaking up matched pairs, one into each drawer, 
knowledge of the content of one drawer will 
also tell you what's in the other.  Hence the ``two 
drawer'' (whole system) uncertainty is reduced to 
$N$ bits.  The added knowledge about how the 
drawers were filled therefore provides $2 N$ 
minus $N$, or $N$ bits of mutual information.  

\subsection{Non-locality} 
\label{subsec: NonLocal}

Where is the mutual information (i.e. those N bits about the 
state of drawers A and B) physically located?  One 
might be tempted to say it's located external to system 
AB, in the observer's database.  In fact, {\em it's 
delocalized} in that changes to either system AB or 
its environment (e.g. a forgetful observer) can change 
things so that knowledge of system A proves 
nothing about system B, i.e. could make the mutual 
information no longer mutual.  Thus mutual information is 
inherently non-local.

\subsection{Correlations \& physical entropy}
\label{subsec:Correlations}

The mutual information itself (i.e. knowledge that the contents 
of the two drawers are correlated) can come either from 
the process by which the 
drawers were filled, or from a peek at their contents.  
Entropy increases in deterministic model physical systems in 
fact depend on the tossing out of correlations between subsystems 
too small to investigate with ``a peek''.  
Thus entropy increases in physical systems can be seen as a loss of 
mutual information between those systems and their environment.  

\begin{figure}
\includegraphics{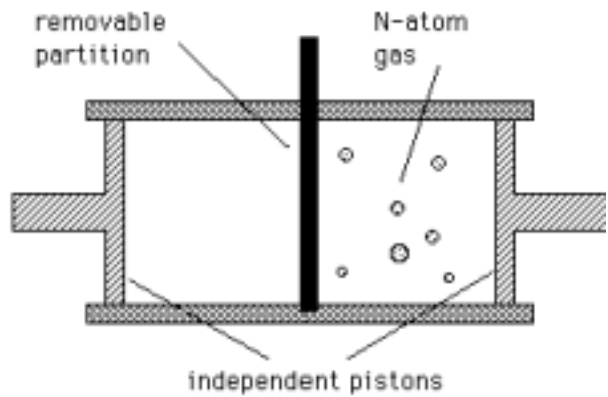}%
\caption{A symmetric bi-partitioned cell for the 
isothermal compression of an N-atom ideal gas into either 
its left or right half, perhaps first discussed by Szilard, 
which serves as a physical system about whose state 
mutual information is available for a well-defined price 
in free energy.  If one is further provided with some 
mechanism (e.g. spectroscopic) for reading its state, it 
may also serve as a mechanically operated single-bit 
memory.  The ``setting'' process involves removing the 
barrier between compartments, using the piston on one side 
to relocate all atoms into the opposite half and then 
returning the barrier before returning the piston to its 
original position.  The required work is $W_{in} = N k T \ln[2]$. 
Its reset status may be defined as true if we know that the 
atoms are located in the right half of the container, and 
false if we don't know this to be the case.}
\label{Szilard}
\end{figure}

For example consider a $1$-atom bi-partitioned 
cell like the $N$-atom cell in Figure \ref{Szilard}.
Using one of the pistons to capture that one ideal gas atom 
on either side of the removable partition decreases the 
entropy of the gas by $k \ln[2]$ or one bit.  If an 
observer is taking notes, it can also increase the mutual 
information between that gas and its environment by one bit 
since they can afterwards answer the true-false question 
correctly:  In which of the two partitions does the atom reside?  

To take a more general view, begin with system A 
having $N$ accessible states so that {\em a priori} uncertainty about A 
is $S_A = k \ln N$.  Then consider an observer B, with sufficient added 
information about A to limit the number of accessible states to $\Omega < N$.  
Observer B therefore has uncertainty about A of $S_{A/B} = k \ln \Omega$.  
What we can learn about A by knowing B also is then the mutual 
information between B and A, i.e. 
$M[A|B] = S_A - S_{A/B}$.  The Second Law assertion that $S_{A/B}$ 
can increase but not decrease with time if systems A and B 
are isolated thus also says quite generally that the mutual information 
between isolated systems A and B can decrease but not increase 
with time.

Even though delocalized mutual information, and 
physical entropy, are connected in this 
way, the latter is historically treated as an extensive 
property ``distributed locally'' throughout the system.  As shown 
by equation \ref{IAB}, uncertainty may be treated as extensive 
i.e. as a sum of parts, as long as subsystem correlations 
within that system (mutual information) may be ignored on the 
macroscopic scale. Then $S_{AB} = S_A + S_B$.  This 
is the case with many thermodynamic systems, 
a limiting case of which are ``ideal gases'' that
behave as though gas molecules are ignoring one another 
completely.

\section{Maxent \& net surprisal}
\label{sec:Maxent}

A robust tool for estimating uncertainties in the 
face of added information might be called ``the 
maxent best-guess machine''.  The added information 
is normally written as ``expected averages'', 
although the approach in principle can accomodate 
a wide range of added information types.  

One first writes entropy in terms of probabilities by defining for 
each probability 
$p_i$ a ``surprisal'' $s_i \equiv k \ln[1/p_i]$, in units determined 
by the value of $k$.  
The average value of this surprisal reduces to $S = k \ln \Omega$ when the $p_i$ are 
all equal, making it simply a logarithmic measure of the effective 
{\em number of accessible states} $\Omega \equiv e^{S/k}$.  The relationships 
described here translate seamlessly into quantum mechanical terms\cite{Jaynes57b}.
 
\subsection{Availability minimization}
\label{subsec:Problem}

The standard problem and its solution has been reiterated 
myriad times since its statement by Jaynes.  The bottom line 
is that the problem of entropy maximization under expectation-value 
constraints can be recast through Lagrange's method of 
undetermined multipliers as a constraint-free {\em minimization} 
of generalized availability in information units.  This 
minimized availability (defined without reference to any 
specific physical system or conservation laws), for example\cite{pf.ifzx}, 
becomes the appropriate free energy for any given thermodynamic ensemble  
when divided by energy's Lagrange multiplier ($1/kT$).

\subsection{Departures \& net surprisal}
\label{subsec:NetSurprisals}

Although availability (hence free energy) minimization 
is quite useful for locating the most likely states, 
finite departures from expected are better measured 
with whole system changes in uncertainty relative to 
a reference state \cite{Evans69, Tribus71}, or {\em net 
surprisals}.  Net surprisal (also called the 
Kullback-Leibler divergence\cite{Kullback51} or relative 
entropy\cite{Badii97}) is defined as
\begin{equation}
I_{net} \equiv - k \sum _{i=1}^{\Omega }{p_i}\ln (\frac{p_{oi}}{p_i}) \geq 0\text{,}
\label{NetSuprisal}
\end{equation}
where the $p_{oi}$ are state probabilities based only 
on ambient state information, while the $p_i$ take into 
account all that is known.  From the solution to the 
standard problem above, one can then show under fairly 
general conditions near ambient\cite{Tribus71, pf.ifzx} that 
derivatives of availability under ensemble 
conditions are also derivatives of net surprisal.

For example, systems in thermal contact with an 
ambient temperature bath may be treated as canonical 
ensemble systems with constrained average energy.  
Thus a temperature deviation from ambient $T_o$ for a 
monatomic ideal gas gives for that system 
$\frac{I_{net}}{k} = \frac{3}{2} N \Xi[\frac{T}{T_o}]$ where 
$\Xi[x] \equiv x - 1 -\ln x \geq 0$.
If that system is also 
in contact with an ambient pressure bath (i.e. able to 
randomly share volume and energy), volume deviations add 
$N \Xi[\frac{V}{V_o}]$ to the foregoing.  For grand canonical 
systems whose molecule types might change (e.g. 
via chemical reaction), one instead adds $N_j \Xi[\frac{N_{jo}}{N_j}]$  
for each molecule type $j$ whose concentration varies 
from ambient\cite{Tribus71, pf.tribus, pf.ifzx}.

  Even more specifically, a problem offered to intro physics students 
at the University of Illinois asked how cool the room must 
be for an otherwise unpowered device to take boiling water 
in at the top, and return it as ice water at the bottom.  
Since the $2^{nd}$ Law allows conversion of one form of 
net surprisal reversibly into another (famously without 
providing any clues how to pull it off), one can simply 
set net surprisals equal for the initial and final states, 
and then solve for $T_{room}$.  

Of course, the net surprisal measure is not only relevant to 
inference about systems for which physically-conserved energy 
is of interest.  For example, it meets the requirements for an information measure 
proposed by Gell-Mann and Lloyd \cite{GellMann96}, and includes 
the Shannon information measure discussed there as a special 
case.  It can also be useful in applied statistics, as in 
the assessment of student responses to multiple choice 
test questions\cite{pf.ifzx}.

\subsection{A special case}
\label{subsec:Mutual_I} 

More importantly, the mutual information 
measure defined earlier is a special case.
To see this, again consider two 
subsystems.  The {\em joint probability} that 
system I is in state i and system J is in state j 
might be written $p[ij]\geq 0$.  This obeys 
$\sum_i \sum_j p[ij] = 1$, where the $i$ indices 
run over all possible states for subsystem I, while the $j$ 
indices run over all possible states for subsystem J.  
From the joint probabilities one can calculate 
{\em marginal probabilities} like $p[i]\equiv \sum_j p[ij]$, 
which ignore the state of other subsystems.  
From these probabilities then values for
{\em joint entropy} $S[IJ]/k\equiv \left<-\ln p[ij] \right>$, and 
{\em marginal entropies} like $S[I]/k\equiv \left<-\ln p[i] \right>$, 
follow immediately.  

Mutual or correlation information between 
systems I and J, denoted here as $M[I|J]$ and defined by 
equation \ref{IAB} as the sum of marginal entropies 
$S[I]+S[J]$ minus the joint entropy $S[IJ]$, thus becomes
\begin{equation}
M[I|J] = - k \sum_i{\sum_j p[ij]}\ln\frac{p[i]p[j]}{p[ij]}\geq 0\text{.}
\label{Mutual2}
\end{equation}
From equation \ref{NetSuprisal}, it is easy to see 
that mutual information is the net surprisal that follows on
learning that two systems (here I and J) are not independent.

One interesting feature of such correlation measures is 
that they refer to the relationship between system A and 
system B, and thus may be quite independent of models 
for system A or B {\em per se}.  One can also express Yaneer Bar-Yam's 
multiscale complexity measures\cite{BarYam04, SGSYBY04} by 
combining sub-system mutual information terms, without reference to 
uncertainty about the state of individual subsystems taken separately.  
For instance with three binary variables 
(e.g. Ising model spins I, J and K), the amount of 
``intermediate scale complexity'' C(2) associated 
with 2 or more spins can be written as
\begin{equation}
M[IJ|K]+M[JK|I]+M[KI|J]-M[I|J|K],
\end {equation}
where $M[I|J|K]$ is the three-system (all-scale) 
mutual information.  Similarly the 
``large scale complexity'' C(3) can be written as
\begin{equation}
2M[I|J|K]-M[IJ|K]-M[JK|I]-M[KI|J].
\end{equation}
These relationships generalize nicely for N spins, 
via the fact that the sum over size-scales from 2 to N 
is the N-system mutual information.  Such mutual 
information expressions also, as discussed above, 
connect such multiscale measures to constraints 
provided by the second law.

  Hence the maxent formalism
allows us to connect the isolated-system 
$2^{nd}$ law to observations like ``If Jimmy 
and Alice didn't talk to each other, there
is no way Jimmy could have known what 
Alice was planning to do.''  This 
assertion is about correlations between 
subsystems (Jimmy and Alice) that are quite 
independent of one's thermodynamic models
for Jimmy and Alice, as are the multiscale 
complexity measures C2 and C3 above.  
Thus armed with statistical inference tools that underpin 
traditional thermodynamic applications, but which require no physical 
assumptions {\em a priori} short of some state inventories, we now take a look 
at some of the more complex system areas where applications (already 
underway in many fields) will likely continue to develop.

Examples of correlated subsystem pairs include 
photon or electron pairs with opposite but 
unknown spins, a single strand of messenger RNA 
and the sequence of nucleotides in the gene from 
which it was copied, a manuscript and a copy of 
that manuscript created with a xerox machine (or 
a video camera), your understanding of a subject 
before being given a test and the answer key 
used by the teacher to grade that test (hopefully), 
enzymes and coenzymes with site specificity, tissue 
sets treated as friendly by your immune system, 
metazoans who developed from the same genetic 
blueprints (e.g. identical twins), families that 
share similar values, and cities which occupy 
similar niches in different cultures (e.g. 
sister cities).  However, does this relationship
between such complex systems and the homogeneous 
systems of physical thermodynamics have any 
consequence in practice?

\section{Correlation physics}
\label{sec:CorrelationPhysics}

\begin{figure}
\includegraphics[scale=0.45]{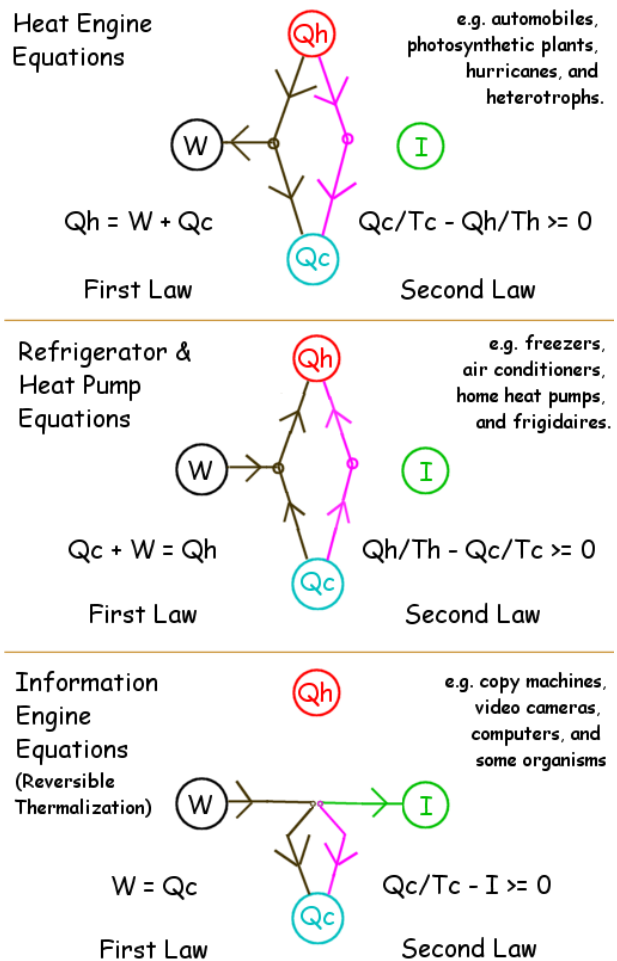}%
\caption{Graphical view of the 
equations underlying some everyday 
thermodynamic engines.}
\label{engines}
\end{figure}

The amount of entropy associated with the flow of 
thermal energy often dwarfs that associated with 
the flow of information, {\em per se}.   For example, the $2^{nd}$ law 
dumps 1/40 eV per nat of erased memory into a 
room temperature ambient, but this is negligible 
compared to other sources of heat in present day 
computers.  Thus thermodynamics is seldom today a 
direct hindrance to information flow.  

\begin{figure*}
\includegraphics[scale=0.6]{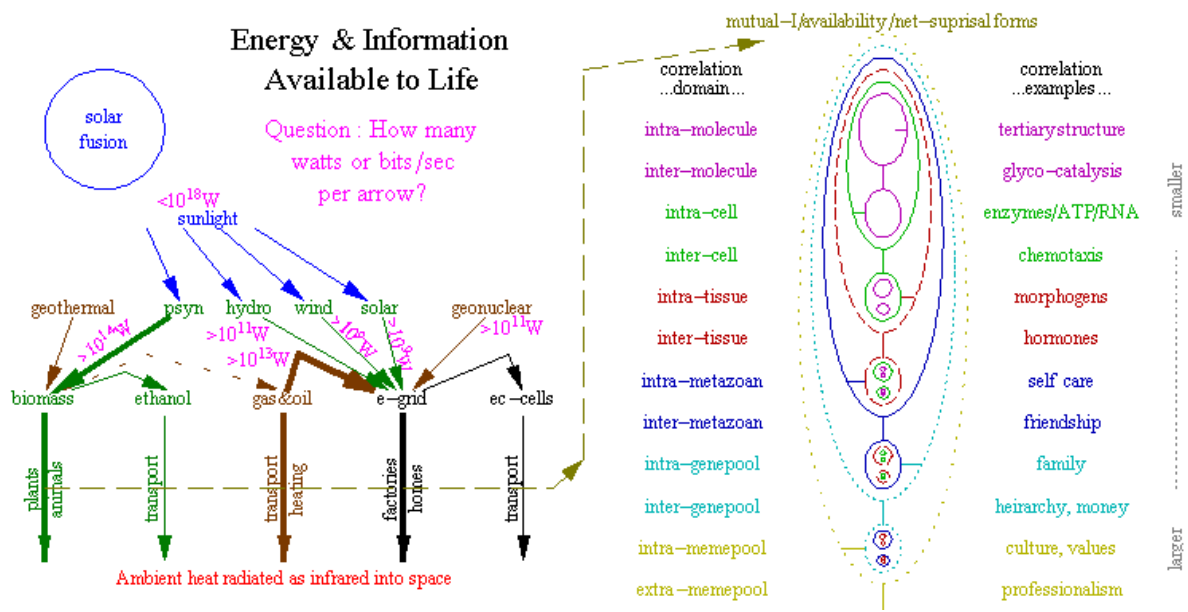}%
\caption{Life's free-energy flow (left), and 
modes of correlation storage (right) powered 
by reversible thermalization of that energy.}
\label{Surprisal}
\end{figure*}

On the other hand, more 
light on a subject (literally free energy to work with) 
rarely hinders discovery.  Perhaps the most striking 
evidence of this practical connection between 
thermal physics and complex systems comes from the 
correlation illustrated by Eric Chaisson, between 
evolved complexity and free-energy flux density\cite{Chaisson04}.  
The latter is a measure of the rate at which 
ordered energy is being thermalized (for the most 
part irreversibly).  The former is evidence, on the 
contrary, that correlations are being effectively 
created and preserved.  

Mutual information (e.g. that two spins are correlated, or 
that two gases have not been well mixed) also plays a well-known 
role in physical systems\cite{Brilloun62, Lloyd89b, Scully03}.  
There is recent focus in particular on its impact 
in nucleic acid replication \cite{Schneider91b, Schneider00} 
and in quantum computing \cite{Wootters82, Gottesman00}.  For example, 
Grosse et al  \cite{Grosse00} use intra-molecule mutual information to 
distinguish coding and non-coding DNA, instead of autocorrelation 
functions, because the former does not require mapping symbols to numbers, 
and because it is sensitive to non-linear and linear dependences.  
Although constraints of this sort may be incorporated into the maxent 
formalism, we take the possibility of such correlations 
into account here by including an internal mutual-information term 
$I_m$ in statements of the isolated system second law, namely  
\begin{equation}
dS = \frac{\delta Q_{in}}{T} - \delta I_m + \delta{S_{irr}} \text{,}
\label{Law2i}
\end{equation}
so that $\delta Q$ becomes $T(dS + \delta I_m - \delta S_{irr})$.  
This makes it possible to consider engines whose primary function 
concerns tasks not explicitly involving changes in energy, 
such as the job of putting ``the kids' socks in one pile, and 
the parents' socks in another''.  

This strategy also reflects work on the energy cost 
of information in generalizing the Maxwell's demon problem \cite{Bennett87}. 
Zurek \cite{Zurek89} among others suggests that the only {\em requisite} 
cost of recording information about other components in a system 
is the cost of preparing the blank sheet (or resetting the measuring 
apparatus) prior to recording with it.  Moreover, the minimum 
thermodynamic cost, in energy per nat of correlation information, 
is simply the ambient temperature $kT$.  A physical 
example\cite{Penrose70} of this is the isothermal compressor 
for an N-atom gas (Fig. \ref{Szilard}) taken for the case when N=1.

Quantitative treatment of correlated physical systems 
also leads naturally to the treatment of engines whose function is to produce  
mutual information or correlations between two systems.  
These correlations might, for example, be 
marble collections sorted by color, a faithful copy of a strand of 
DNA, or dots on a sky map corresponding to the position of stars in 
the night sky.  Familiar first and second law engine equations 
then become
\begin{equation}
dU = (\delta Q_{out}) - (\delta W_{in}) = 0\text{, and}
\label{IELaw1}
\end{equation}
\begin{equation}
dS = \frac{\delta Q_{out}}{T_{out}} - \delta I_m = \delta S_{irr} \geq 0\text{.}
\label{IELaw2}
\end{equation}
Eliminating $Q_{out}$ from these two equations yields 
\begin{equation}
\delta I_m \leq \frac{\delta W_{in}}{T}\text{.}
\label{IEresult}
\end{equation}
This means that information engines can produce no more mutual 
information than their energy consumption, divided by 
their ambient operating temperature.  In binary information units, 
this amounts to producing about 55 bits of information per eV of 
thermalized work at room temperature, and around 60 bits per eV 
of energy if operating near the freezing point of water.  
The equations for these engines are compared graphically 
to those of heat engines and heat pumps in Figure \ref{engines}.

Cameras, tape recorders, and copying machines may be considered 
such information engines, as are forms of life that take in 
chemical energy available for work from plant biomass, and thermalize 
that energy at ambient temperature while creating correlations 
between objects in their environment and their survival needs, 
or in the form of informed DNA sequences, 
songs, rituals, books, etc.  This may in fact be a key role 
for living organisms which are not primary producers\cite{Odum71}.

The relevance of these concepts to sustainability 
is illustrated in Figure \ref{Surprisal}.  
The top left illustrates the primary processes supplying 
free energy, while the bottom left 
illustrates repositories as well as paths for thermalizing 
that energy for eventual ambient radiation into space.  
The right half of Fig. \ref{Surprisal} 
tracks correlations that are created (to the extent 
that this thermalization is reversible), 
and categorizes them with (horizontal/vertical) bars 
representing correlations looking (in/out)-ward with respect
to the physical boundary that they relate.  
This story of emergent correlations, as a fringe benefit of 
otherwise irreversible process, fits
beautifully into a much larger picture. 

\section{Inventions, Excitations \& Codes}
\label{sec:CodesExcitations}

Begin with a hierarchical look at the evolution of 
complexity in time\cite{Chaisson04, pf.tribus}.
An abstract and partial outline of the result is provided 
in Table \ref{lodetail}, which also attempts to 
highlight the pivotal role of spatially-defined 
boundaries and gradients in the natural history of 
invention.

\begin{table}
\caption{``Temporally-stacked'' layers of correlation-based 
complexity.}
\begin{tabular}{r|c|l}
new drivers 			& boundaries 			& emergence 	\\ 
\hline \hline
stable 						& voltage 				& neutral 		\\
nuclei 						& gradients 			& matter 			\\
\hline
density 					& gravity 				& forming 		\\
fluctuations 			& gradients 			&	galaxies		\\
\hline
interstellar 			& temperature 			& stellar 		\\
clouds  					& gradients 			& ignition		\\
\hline
orbiting  				& radial pressure 	& planets with \\ 
dust \& gas 			& variation 			& geocycles 	\\ 
\hline
geothermal \& 			& compositional 		& biomolecule 	\\ 
solar energy 			& variation 			& cycles 			\\ 
\hline
biological  			& bilayer 				& microbial 	\\ 
cells						& walls						& symbioses 	\\ 
\hline
biofilms \& 			& organ  					& multi-organ 	\\
live tissues 			& surfaces 				& systems 		\\  
\hline
metazoans 				& individual 			& pair bonds,	\\
								& skins 					& redirection 	\\
\hline
reproductive 			& gene-pool 			& hierarchies 	\\ 
bargains, family 	& boundaries 			& \& money		\\ 
\hline
cultures \&  			& meme-pool 				& sciences \& 	\\ 
belief systems 		& boundaries 			& ``diversity'' \\ 
\hline \hline
\end{tabular}
\label{lodetail}
\end{table}

The standing crop of correlations in each of these 
cases involves physical boundaries of increasing 
complexity.  These range from gradients (e.g. of 
temperature, pressure, or composition), through 
diffusion boundaries (e.g. bi-layer membranes
through metazoan skins), to gene and meme pool 
boundaries which are fiendishly complex but 
physical nonetheless.  Associated with each 
boundary or gradient are also availability fluxes, 
like neutralizing charge transported over 
voltage gradients, PdV work done in crossing pressure 
gradients, nutrients through cell walls, 
blood flow between tissue systems, and territorial 
flows between families.

The approach also allows one to follow Chaisson's 
lead and discuss the emergence of complex systems 
in an integrated context.  One advantage of this is that 
we have ``live observation'' of stars, planets, and weather 
emerging in many places, even if we don't yet have other 
examples e.g. of biomolecule cycles giving rise 
to membrane-protected cells.  Detailed ``timelines of 
concept-relevance'' strengthen this integrative picture, 
since concepts of ordered energy and mutual information repeatedly 
intertwine in non-repeating but self-similar 
fashion\cite{pf.ifzx}.  

An example of this self-similarity is the 
invention of money as a ritualized reminder of 
expended available work.  Another is the emergence 
of replicable molecular codes complementary to 
the survival of cell groups, just as replicable ideas in
human society play a role complementary to the
survival of groups of individuals.  

\section{Monitoring correlations}
\label{sec:Monitoring}

\subsection{Inventories of standing crop}

Perhaps the simplest thing to do is count. Doing this objectively in practice, of course, is far from trivial. Further, more detailed consideration (e.g. ennumeration of alternatives) may be necessary to put such measures into $2^{nd}$ law terms. Nonetheless, one might sketch the outline of correlation measures in the following way across three quite different levels of complex system organization: the state of a molecule, a metazoan, and a community of individuals.

\begin{itemize}
\item (A)  $N_{correlated states/fermion} \times  N_{fermions/molecule}$,
\item (B) $N_{assignments/molecule} \times N_{molecules/metazoan}$,
\item (C) $N_{niches/metazoan} \times N_{metazoans/community}$.
\end{itemize}

Presumably the first term in each case has an upper limit.  
For example, assume that individual elements each occupy 
no more than one correlated state directed inward, and/or 
outward, from each of the physical boundary types 
that comprises the level of which they are part.  

Thus with a community of individuals we might
conceptualize niches as focussed inward and outward with respect 
to the boundaries of self (physical skin), family (gene-pool), and 
culture (meme-pool).  Although the latter two boundaries 
between groups of correlated codes are geometrically complicated 
(to say the least), they are {\em physical} boundaries nonetheless.  
Roles taking care of self (skin IN), friends (skin OUT), 
family (gene-pool IN), hierarchy (gene-pool OUT), 
culture (meme-pool IN), and profession (meme-pool OUT) have 
thus developed, as have (respectively) the 
related lore and participant/leadership obligations of 
patient/doctor, colleague/mentor, sibling/parent, 
citizen/leader, dancer/priest, and professional/scholar. 
In other words, this inventory for the case of a community 
simply asks: ``In how many of these six areas are individuals, 
on average, fortunate enough to be able to make a name for 
themselves?'' If this is decreasing, things are perhaps getting 
worse for the community, independent of what other indicators 
have to say.

On the level of metazoan, we might similarly consider molecule assignments 
pointing inward and outward with respect to molecule surface, cell-membrane, 
and organ.  For example, hormone molecules required to 
convey signals from one organ system to another might 
be seen as charged with an inter-tissue (organ OUT) assignment.  

On the level of an individual molecule, the relevant 
boundaries for correlated fermion states might be the fermion wave-packet, the 
appropriate nuclear or electronic shell, and the atom comprising the molecule.  
Electrons involved in co-valent bonds might in this sense be involved in correlations 
directed outward from the atom to which they were initially assigned.  
Internal to atoms, counting correlations in second-law (mutual information) 
terms may be easier still.  For example, the mutual information 
between up-spin and down-spin electrons comprising a He atom's K-shell 
is simply one bit.  

In the expressions above, the number of states/assignments/niches per agent is an average, 
so that each total can also be determined by a sum of all the states/assignments/niches 
in the larger unit. Each of the three bullets above, in sequence, attempts to estimate the 
correlation-information associated with order on a larger size scale. Thus 
each assignment of a metazoan molecule builds on a certain number of correlated 
fermion-states within that molecule, just as each niche for a community metazoan will 
build on a certain number of correlated molecule-assignments within that metazoan. 
Nonetheless each assignment or niche, as an emergent phenomenon, is quite distinct 
from (hence something more than) the sum of its constituent states or assignments, 
respectively.

\subsection{Process indicators}

Ultimately, the goal is to look at the rates at which the correlations above are created, minus the rates at which they are lost. One strategy, of course, might be to track the inventories above as a function of time. Those measures, at least, will presumably be consistent with other measures of these rates.

Thanks to the $2^{nd}$ law of thermodynamics, we also have the fact that rates of reversible thermalization are less than or equal to the rates of available work used up. Thus the rate at which available work is fed into the system (hence at which something is lost in the figure above) offers an upper limit on the rate (when converted into second law terms) at which correlations are gained. Such energy-based measures include life's power stream here on earth (left side of Fig. \ref{Surprisal}), and more generally Eric Chaisson's ``free energy rate density''\cite{Chaisson98, Chaisson04}.

\section{Synthesis \& Conclusions}
\label{sec:Conclude}

Here we've discussed net surprisal 
tools that can be applied to mutual
information between parts of more complex systems, 
independent of the existence of models for 
individual subsystems per se.  Such applications 
were conspicuously launched by Shannon in the 
late 1940's.  Although their application in everyday 
and cutting-edge applications is expanding today 
at an even faster rate, awareness of a cross-disciplinary 
foundation for applying them remains incomplete.  

We also discuss the utility of net surprisals 
for treating reversible equilibration problems 
(like a device for reversibly converting hot
water into cold), and suggest 
that physical boundaries in complex systems (like 
code-pool boundaries) provide a platform for 
the quantitative monitoring of subsystem correlations 
in systems as complex as human communities today. 
The hard work of putting these observations to 
use is ahead.

\begin{acknowledgments}
I would like to thank the late E. T. Jaynes for notes on entropy 
maximization and the binomial distribution, as well as for 
many interesting papers and discussions over the past half 
century, Myron Tribus for inspiration and some challenging 
questions, and Tom Schneider at NIH for spirited 
questions and comments as well.
\end{acknowledgments}



\bibliography{ifzx}

\end{document}